\documentclass[aps,prc,preprint,a4paper,groupedaddress,showpacs,amsmath]{revtex4-1}

\usepackage{graphicx}
\usepackage{dcolumn}
\usepackage{amsmath}
\usepackage{amssymb}

\begin{document}

\bibliographystyle{prsty}

\title{Role of momentum transfer in the quenching of Gamow-Teller strength}

\author{T. Marketin}
\affiliation{Institut f\"{u}r Kernphysik, Technische Universit\"{a}t Darmstadt, D-64289 Darmstadt, Germany}
\affiliation{Physics Department, Faculty of Science, University of Zagreb, 10000 Zagreb, Croatia}
\author{G. Mart\'{i}nez-Pinedo}
\affiliation{Institut f\"{u}r Kernphysik, Technische Universit\"{a}t Darmstadt, D-64289 Darmstadt, Germany}
\author{N. Paar}
\affiliation{Physics Department, Faculty of Science, University of Zagreb, 10000 Zagreb, Croatia}
\author{D. Vretenar}
\affiliation{Physics Department, Faculty of Science, University of Zagreb, 10000 Zagreb, Croatia}

\date{\today}

\begin{abstract}
\begin{description}
\item[Background] Differential cross sections for the $(p,n)$ and $(n,p)$ reactions on $^{90}$Zr over the interval of 
$0 - 50$ MeV excitation energy were used to determine the corresponding GT strengths, and the resulting 
quenching factor $\approx 0.9$ with respect to the Ikeda sum rule. In this procedure the contribution of the isovector spin monopole (IVSM) strength was subtracted from the total strength without taking into account the interference between the GT and the IVSM modes. 
\item[Purpose] To determine the quantitative effect of the IVSM excitation mode on the $L=0$ strength in charge-exchange reactions on several closed-shell nuclei and the Sn isotopic chain. 
\item[Method] The fully consistent relativistic Hartree-Bogoliubov (RHB) model + proton-neutron quasiparticle random phase approximation (pn-RQRPA) is employed in the calculation of transition strength in the 
$\beta^{-}$ and $\beta^{+}$ channels.
\item[Results] The inclusion of the higher-order terms, that include the effect of finite momentum transfer, in the transition operator shifts a portion of the strength to the high-energy region above the GT resonance. The total strength is slightly enhanced in nuclei with small neutron-to-proton ratio but remains unchanged with increasing neutron excess.
\item[Conclusions] Terms that include momentum transfer in the transition operator act mostly to shift the 
strength to high excitation energies, but hardly affect the total strength. Based on the strength obtained using the full $L=0$ transition operator in the pn-RQRPA calculation, we have estimated the impact of the IVSM on the strength measured in the charge-exchagne reactions on $^{90}$Zr and found that the data are consistent with the Ikeda sum rule. 

\end{description}
\end{abstract}


\maketitle
\section{Introduction}

Collective spin and isospin excitations in atomic nuclei have been the subject of many experimental and theoretical studies (for an extensive review see Ref.~\cite{Osterfeld1992}). Of particular interest is the Gamow-Teller (GT) resonance, a collective oscillation of neutrons that coherently change the direction of their spins and isospins without changing their orbital motion. A detailed knowledge of the GT strength distribution is essential for the understanding of nuclear beta-decay and weak processes in stars~\cite{Langanke2003}. It was first predicted in 1963~\cite{Ikeda1963} and observed in $(p,n)$ reactions a decade later~\cite{Doering1975}.  Further measurements identified a problem that is still actively discussed, and that is the quenching of the Gamow-Teller strength. In numerous experiments across the whole nuclear chart only around 60\% of the strength predicted by the model-independent Ikeda sum rule had been observed~\cite{Osterfeld1992}. These experiments, however, were only able to measure the strength up to the excitation energy of the giant resonance. A consistent analysis of $(p,n)$ and $(n,p)$ reaction data from $^{90}$Zr over a much wider range of excitation energies, concluded that the GT strength is actually quenched by approximately 10\%~\cite{Wakasa1997,Yako2005}. Theoretical models systematically overestimate the transition strength compared to the measured values~\cite{Caurier2005}. This effect was attributed to two possible processes: (i) coupling of the GT mode to $\Delta$-isobar nucleon-hole ($\Delta - h$) configurations, and (ii) second-order configuration mixing. It has been shown that the former process is responsible for only a small fraction of the quenching~\cite{Ichimura2006,Towner1987}, leaving the latter as the major mechanism for shifting the GT strength to higher energies~\cite{Drozdz1986a}. 

The  spin-isospin operator structure of the $(p,n)$ probe is similar to that of the Gamow-Teller (GT) operator~\cite{Ichimura2006}. However, they become comparable only if the GT cross section is measured at very small momentum transfer $q$. In the $(p,n)$ reaction this condition can be met only for zero degree scattering, small excitation energies, and high bombarding energies. Extraction of the $L=0$ strength at high excitation energies, where higher-multipole response dominates, is very difficult~\cite{Vetterli1989,Anderson1990}. Nevertheless, recent experiments reported data on the Gamow-Teller response in $^{90}$Zr, in both $\beta^{-}$~\cite{Wakasa1997} and $\beta^{+}$ channels~\cite{Yako2005}, up to 50 MeV excitation energies. Therefore, in the total strength, contributions from higher-order terms in the expansion in $q$ appear, the first of which is the isovector spin monopole (IVSM) mode. This mode, with the transition operator $r^{2}\boldsymbol{\sigma}\tau$, represents a collective excitation of the nucleus with quantum numbers $J^{\pi}=1^{+}$, $L=0$, $S=1$ and $T=1$. Even though the first observation of the isovector spin monopole mode was reported in 1983~\cite{Bowman1983}, a quantitative analysis and determination of the strength remain difficult~\cite{Prout2000,Zegers2001,Zegers2004}. The unknown IVSM strength also introduces uncertainties in the measurement of the Gamow-Teller strength. To obtain precise quantitative data on the total GT strength, and indirectly on the quenching, the contribution of the isovector spin monopole mode and of higher-order terms in momentum transfer, must be subtracted from the measured strength. Although interference occurs between the GT and IVSM modes, these contributions are usually subtracted incoherently from the spectrum because the distribution of GT strength in the IVSM resonance region is unknown~\cite{Yako2005}. Related to new measurements of nuclear response in unstable nuclei~\cite{Sasano2011}, the correct treatment of effects that influence the extraction of the Gamow-Teller strength is all the more important.

Not much theoretical work has been reported on the IVSM strength so far. Isovector spin excitations with angular momentum $L=0,1$, and $2$ have been studied in Ref.~\cite{Auerbach1984a} employing the Skyrme SIII 
Hartree-Fock model and the random-phase approximation (RPA) with a schematic residual p-h interaction. 
The Skyrme functionals SGII and SIII  were used in a self-consistent HF + Tamm-Dancoff approximation (TDA) study that focused only on the IVSM mode in $^{48}$Ca, $^{90}$Zr and $^{208}$Pb~\cite{Hamamoto2000}. In both cases the IVSM mode was identified at excitation energies between 20 MeV and 60 MeV with respect to the ground state energy of the parent nucleus. A non-energy-weighted sum rule was devised that involves particle numbers and radii, and the calculated strengths were shown to be consistent with values obtained from ground-state densities. 
The effect of the isovector spin monopole mode and higher-order terms in momentum transfer on the quenching of GT strength has not been investigated so far. 

In this study we explore the IVSM mode of excitation within a self-consistent microscopic theory, and analyze the effect of momentum transfer on spin-isospin excitations. The relativistic Hartree-Bogoliubov (RHB) + proton-neutron relativistic quasiparticle random phase approximation (pn-RQRPA) framework is employed for the calculation of the IVSM strength. This framework, based on the covariant energy density functional theory, is a charge-exchange extension of the relativistic quasiparticle RPA formulated in the canonical basis of the RHB model~\cite{Paar2003}. The RHB + pn-RQRPA have already been successfully applied to the analysis of the Fermi and the Gamow-Teller response~\cite{Paar2004}, $\beta$-decay half-lives~\cite{Niksic2005,Marketin2007}, neutrino-nucleus cross-sections~\cite{Paar2008}, total muon capture rates~\cite{Marketin2009}, and electron capture rates~\cite{Niu2011}.

The IVSM strength is calculated for the closed-shell nuclei $^{48}$Ca, $^{90}$Zr and $^{208}$Pb. The isovector spin monopole operator is also considered in the context of an expansion of the transition operator with respect to the momentum transfer in the reaction. The impact of the momentum transfer on the total $L=0$ strength and its distribution is examined in the Sn isotopic chain between $A=100$ and $A=150$. Sec. \ref{sec:theory} introduces the formalism, and Sec. \ref{sec:results} presents the results and discussion. Sec. \ref{sec:conclusion} contains a short summary and concluding remarks.

\section{Theoretical formalism} \label{sec:theory}

The relativistic quasiparticle random phase approximation (RQRPA) was formulated in the canonical single-nucleon basis of the RHB model in Ref.~\cite{Paar2003} and extended to the description of charge-exchange excitations (pn-RQRPA) in Ref.~\cite{Paar2004}. The RHB + RQRPA model is fully self-consistent: in the particle-hole channel, effective Lagrangians with density-dependent meson-nucleon couplings are employed, and pairing correlations are described by the pairing part of the finite range Gogny interaction~\cite{Berger1991}. In both the \emph{ph} and \emph{pp} channels the same interactions are used in the RHB equations that determine the canonical quasiparticle basis, and in the matrix equations of the RQRPA. This is very important because the energy weighted sum rules are fulfilled only if the pairing interaction is consistently included both in the static RHB and in the dynamical RQRPA calculation. In the present work all calculations are performed using one of the most accurate meson-exchange density-dependent relativistic mean-field interactions in the \emph{ph} channel: DD-ME2~\cite{Lalazissis2005}.

Transitions between the $0^+$ ground state of a spherical even-even parent nucleus and the $J^\pi$ excited state of the corresponding odd-odd daughter nucleus are induced by a charge-exchange operator $T^{JM}$. Taking into account the rotational invariance of the nuclear system, the quasiparticle pairs are coupled to good angular momentum and the matrix equations of the pn-RQRPA read:
\begin{equation} \left( \begin{array} [c]{cc}
A^{J} & B^{J}\\
B^{^{\ast}J} & A^{^{\ast}J}
\end{array}
\right)  \left( \begin{array} [c]{c}
X^{\lambda J}\\
Y^{\lambda J}
\end{array}
\right)  =E_{\lambda}\left( \begin{array} [c]{cc}
1 & 0\\
0 & -1
\end{array}
\right)  \left( \begin{array} [c]{c}
X^{\lambda J}\\
Y^{\lambda J}
\end{array}\right) \; . 
\label{eq:pnrqrpaeq}
\end{equation}
The matrices $A$ and $B$ are defined in the canonical basis~\cite{Ring1980}
\begin{eqnarray}
A_{pn,p^\prime n^\prime}^{J} &=& H^{11}_{pp^\prime}\delta_{nn^\prime} +
  H^{11}_{nn^\prime}\delta_{pp^\prime}  \nonumber \\ & & +
\left( u_p v_n u_{p^\prime} v_{n^\prime} + v_p u_n v_{p^\prime} u_{n^\prime}\right)
 V_{pn^\prime n p^\prime}^{ph J} + 
\left( u_p u_n u_{p^\prime} u_{n^\prime} + v_p v_n v_{p^\prime} v_{n^\prime}\right) 
 V_{pn p^\prime n^\prime}^{pp J} \nonumber \\
B_{pn,p^\prime n^\prime}^{J} &=& (-1)^{j_{p^\prime}-j_{n^\prime}+J}
\left( u_p v_n v_{p^\prime} u_{n^\prime} + v_p u_n u_{p^\prime} v_{n^\prime}\right)
 V_{pp^\prime n n^\prime}^{ph J} \nonumber \\ & &- 
\left( u_p u_n v_{p^\prime} v_{n^\prime} + v_p v_n u_{p^\prime} u_{n^\prime}\right) 
 V_{pn p^\prime n^\prime}^{pp J} \; .
\label{eq:abmat}
\end{eqnarray}
Here $p$, $p^\prime$, and $n$, $n^\prime$ denote proton and neutron quasiparticle canonical states, respectively, $V^{ph}$ is the proton-neutron particle-hole residual interaction, and $V^{pp}$ is the corresponding particle-particle interaction. The canonical basis diagonalizes the density matrix, and the occupation amplitudes $v_{p,n}$ are the corresponding eigenvalues.  However, the canonical basis does not diagonalize the Dirac single-nucleon mean-field Hamiltonian $\hat{h}_{D}$ and the pairing field $\hat{\Delta}$, and therefore the off-diagonal matrix elements $H^{11}_{nn^\prime}$ and $H^{11}_{pp^\prime}$ appear in Eq. (\ref{eq:abmat}):
\begin{equation} \label{H11}
H_{\kappa \kappa^\prime}^{11}=(u_{\kappa }u_{\kappa^\prime } -v_{\kappa }v_{\kappa^\prime })h_{\kappa \kappa^\prime }-(u_{\kappa }v_{\kappa^\prime } + v_{\kappa }u_{\kappa^\prime })\Delta _{\kappa \kappa^\prime }\;,
\end{equation}
For each energy $E_{\lambda}$, $X^{\lambda J}$ and $Y^{\lambda J}$ in Eq. (\ref{eq:pnrqrpaeq}) denote the corresponding forward- and backward-going QRPA amplitudes, respectively. The total strength for the transition between the ground state of the even-even (N,Z) nucleus and the excited state of the odd-odd (N+1,Z-1) or (N-1,Z+1) nucleus, induced by the operator $T^{JM}$, reads
\begin{equation} \label{eq:strength-}
B_{\lambda J}^{\pm} = \left| \sum_{pn} <p||T^J||n> \left( X_{pn}^{\lambda J} u_p v_n + (-1)^J Y_{pn}^{\lambda J}v_p u_n \right) \right|^2 \; .
\end{equation}
The discrete strength distribution is folded by the Lorentzian function
\begin{equation}
R(E)^{\pm} = \sum_{\lambda} B_{\lambda J}^{\pm} \frac{1}{\pi} \frac{\frac{\Gamma}{2}}{\left( E - E_{\lambda_{\pm}}\right)^{2} + \left( \frac{\Gamma}{2} \right)^{2}}.
\end{equation}
In the present calculation the width of the Lorentzian function is $\Gamma = 1$ MeV.

The spin-isospin interaction terms are generated by $\rho$-and $\pi$-meson exchange. Because of parity conservation, the one-pion direct contribution vanishes in the mean-field calculation of a nuclear ground state. Its inclusion is important, however, in calculations of excitations that involve spin and isospin degrees of freedom. The particle-hole residual interaction in the pn-RQRPA is derived from the Lagrangian density 
\begin{equation}
\mathcal{L}_{\pi + \rho}^{int} = 
      - g_\rho \bar{\psi}\gamma^{\mu}\vec{\rho}_\mu \vec{\tau} \psi 
      - \frac{f_\pi}{m_\pi}\bar{\psi}\gamma_5\gamma^{\mu}\partial_{\mu}
        \vec{\pi}\vec{\tau} \psi \; . 
\label{eq:lagrres}	
\end{equation}
Vectors in isospin space are denoted by arrows, and boldface symbols indicate vectors in ordinary three-dimensional space.

The coupling between the $\rho$-meson and the nucleon is assumed to be a function of the vector density $\rho_{v} = \sqrt{j_\mu j^\mu}$, with $j_{\mu} = \bar{\psi}\gamma_\mu \psi$. In Ref.~\cite{Niksic2002a} it has been shown that the explicit density dependence of the meson-nucleon couplings introduces additional rearrangement terms in the residual two-body interaction of the RRPA, and that their contribution is essential for a quantitative description of excited states. However, since the rearrangement terms include the corresponding isoscalar ground-state densities, it is easy to see that they are absent in the charge exchange channel. 
For the $\rho$-meson coupling the functional form used in the DD-ME2 density-dependent effective interaction~\cite{Lalazissis2005} reads
\begin{equation}
g_{\rho} (\rho_{v}) = g_{\rho}(\rho_{sat}) e^{-a_{\rho} (x-1)}\;,
\end{equation}
where $x=\rho_{v} / \rho_{sat}$, and $\rho_{sat}$ denotes the saturation vector density in symmetric nuclear matter. For the pseudovector pion-nucleon coupling the standard parameters are used (see Ref. \cite{Walecka2004}),
\begin{equation}
m_{\pi}=138.0~\text{MeV}~~~~\;\;\;\;\frac{\;f_{\pi}^{2}}{4\pi}=0.08\;.
\end{equation}
The derivative type of the pion-nucleon coupling necessitates the inclusion of a zero-range Landau-Migdal term, which accounts for the contact part of the nucleon-nucleon interaction
\begin{equation}
V_{\delta\pi} = g^\prime \left( \frac{f_\pi}{m_\pi} \right)^2 \vec{\tau}_{1} \vec{\tau}_{2} \boldsymbol{\Sigma}_{1} \cdot \boldsymbol{\Sigma}_{2} \delta (\boldsymbol{r}_{1}-\boldsymbol{r}_{2})\; ,
\label{eq:deltapi}
\end{equation}
where
\begin{equation}
\boldsymbol{\Sigma} = \left( \begin{array}{cc}
\boldsymbol{\sigma} & 0 \\
0 & \boldsymbol{\sigma} \end{array} \right),
\end{equation} 
and the parameter $g_{DD-ME2}' = 0.52$ is adjusted to reproduce the GTR excitation energy in $^{208}$Pb.

The pn-RQRPA model is fully consistent: the same interactions, both in the particle-hole and particle-particle channels, are used in the RHB equation that determines the canonical quasiparticle basis, and in the pn-RQRPA Eq. (\ref{eq:pnrqrpaeq}). In both channels the same strength parameters of the interactions are used in the RHB and RQRPA calculations.  With respect to the RHB calculation of the ground state of an even-even nucleus, 
the charge-exchange channel includes the additional one-pion exchange contribution.

The two-quasiparticle configuration space includes states with both nucleons in the discrete bound levels, states with one bound nucleon and one nucleon in the continuum, and also states with both nucleons in the continuum. In addition to configurations built from two-quasiparticle states of positive energy, the RQRPA configuration space contains pair-configurations formed from fully or partially occupied states of positive energy and empty negative-energy states from the Dirac sea. The inclusion of configurations built from occupied positive-energy states and empty negative-energy states is essential for the consistency of the model~\cite{Paar2004}.

In the $pp$-channel of the RHB model a phenomenological pairing interaction is used, the pairing part of the Gogny force,
\begin{equation}
V^{pp}(1,2)~=~\sum_{i=1,2}e^{-((\mathbf{r}_{1}-\mathbf{r}_{2})/{\mu _{i}}%
)^{2}}\,(W_{i}~+~B_{i}P^{\sigma }-H_{i}P^{\tau }-M_{i}P^{\sigma }P^{\tau }),
\label{eq:Gogny}
\end{equation}
with the set D1S \cite{Berger1984} for the parameters $\mu _{i}$, $W_{i}$, $B_{i} $, $H_{i}$ and $M_{i}$ $(i=1,2)$. This force has been very carefully adjusted to the pairing properties of finite nuclei all over the periodic table. In particular, the basic advantage of the Gogny force is the finite range, which automatically guarantees a proper cut-off in the momentum space. The same Gogny interaction is also used in the $T=1$ pairing channel of the pn-RQRPA.
 
\section{ \label{sec:results} Results}

In the first part the calculated strength distributions for the isovector spin monopole transition operator in 
$^{90}$Zr and $^{100-150}$Sn are analyzed. We then calculate the $L=0$ strength and show the effect of  momentum transfer on the strength distribution and the total strength in the tin isotopic chain. 
\subsection{ \label{sec:ivsm} Isovector spin monopole strength}

The isovector spin monopole (IVSM) operator reads
\begin{equation} \label{eq:ivsmop}
T_{\pm}^{IVSM} = \sum_{i=1}^{A} r_{i}^{2} \boldsymbol{\Sigma} \tau_{\pm}\; .
\end{equation}

\begin{figure}
\centerline{ 
  \includegraphics[scale=0.65]{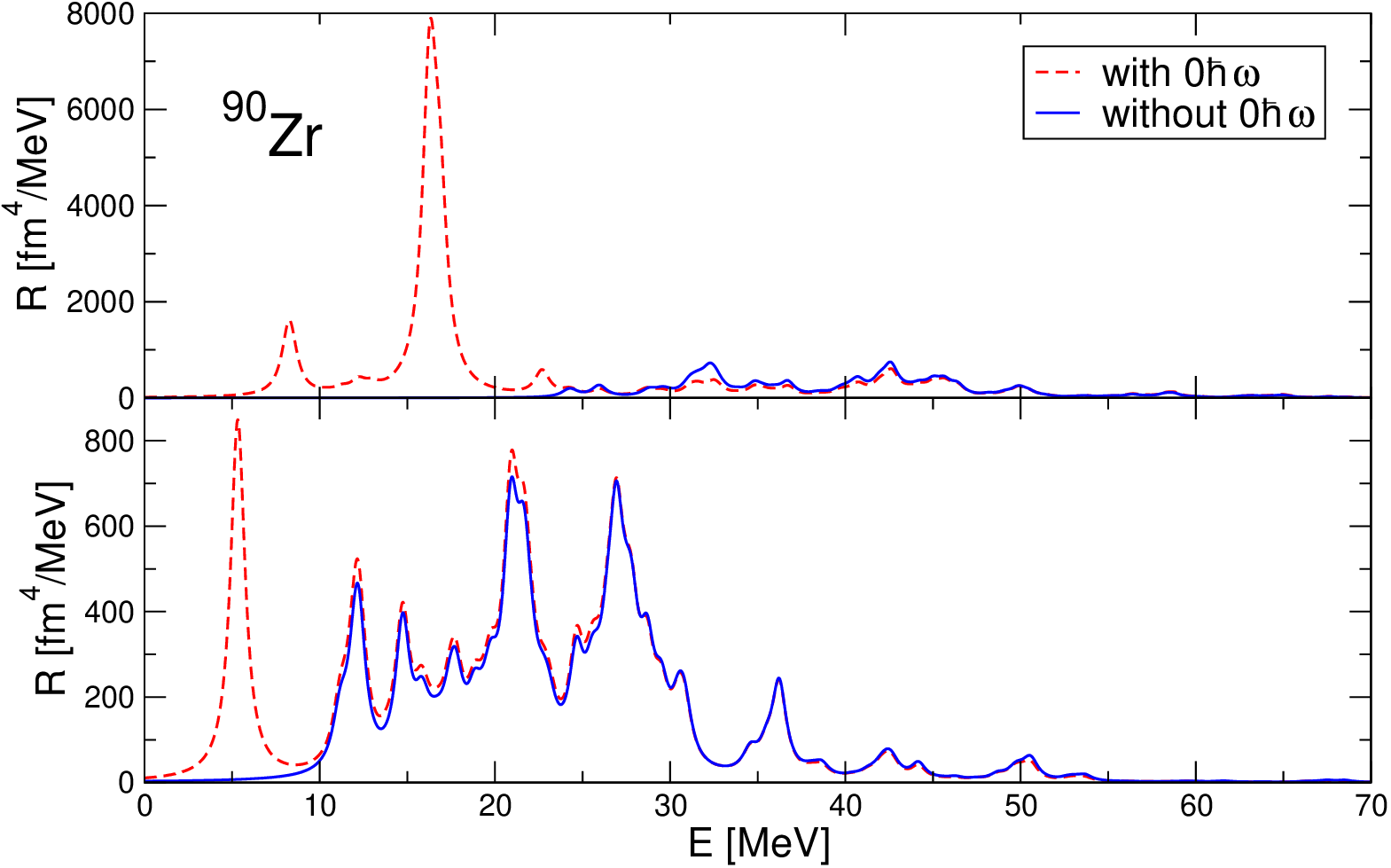}%
}
\caption{\label{fig:ivsmstrength} (color online) The pn-RQRPA strength distribution of isovector spin monopole states in $^{90}$Zr. The dashed curve denotes the total strength with $0\hbar\omega$ transitions included, whereas the solid curve corresponds to the strength for which only $2\hbar\omega$ and higher excitations are included in the configuration space, for the $\beta^{-}$ and the $\beta^{+}$ channels in the upper and lower panels, respectively.}
\end{figure}

In Fig.~\ref{fig:ivsmstrength} we display the IVSM strength for $^{90}$Zr up to 70 MeV excitation energy.
Because of the structure of the IVSM operator of Eq. (\ref{eq:ivsmop}), the dominant feature of the spectrum is a strong peak at the position of the Gamow-Teller (GT) resonance. However, unlike the GT operator that excites only $0\hbar \omega$ transitions, the isovector spin monopole operator can also excite $2\hbar \omega$ transitions. These relatively weak transitions at excitation energies above the giant resonance contribute a significant portion of the total strength, because even though each individual transition is weak, their number is large. The $2\hbar\omega$ transitions are also responsible for all the strength above the GT resonance in the $\beta^{+}$ channel. The amount of strength above the resonance in the two channels is comparable, so that the value of the sum rule is mostly determined by the strength contained in the resonances. 

The non-energy-weighted sum rule for the IVSM transition strength reads~\cite{Auerbach1984a}
\begin{equation} \label{eq:ivsmsumrule}
S_{-} - S_{+} = 3 \left[ N \left\langle r^{4} \right\rangle_{n} - Z \left\langle r^{4} \right\rangle_{p} \right].
\end{equation}
Using the values of $\left\langle r^{4} \right\rangle_{n,p}$ that correspond to  the RHB self-consistent ground-state solution, we have verified that the sum rule Eq. (\ref{eq:ivsmsumrule}) is satisfied in our calculation (cf. Table \ref{tab:ivsmsumrule}).
\begin{table}[hbtp]
\caption{ \label{tab:ivsmsumrule} Integrated strengths of the isovector spin monopole operator in $^{48}$Ca, $^{90}$Zr and $^{208}$Pb nuclei. Proton and neutron radii correspond to the RHB self-consistent ground solution. All values are given in units of fm$^{4}$.}
\begin{ruledtabular}
\begin{tabular}{c|cccccc}
 & $S_{-}$ & $S_{+}$ & $S_{-} - S_{+}$ & $\left\langle r^{4} \right\rangle_{n}$ & $\left\langle r^{4} \right\rangle_{p}$ & $3[N \left\langle r^{4} \right\rangle_{n} - Z \left\langle r^{4} \right\rangle_{p}]$ \\
\hline
$^{48}$Ca & 11037 & 2837 & 8200 & 227.382 & 181.359 & 8218 \\
$^{90}$Zr & 28773 & 11772 & 17001 & 438.375 & 409.187 & 16654 \\
$^{208}$Pb & 266890 & 43795 & 223095 & 1318.019 & 1118.283 & 223095 \\
\end{tabular}
\end{ruledtabular}
\end{table}

Since the IVSM strength distribution is dominated by the $0\hbar\omega$ components, to study the behavior of the IVSM mode 
it is convenient to exclude these transitions. This can be done by including only $2\hbar \omega$ and higher configurations in the QRPA basis. The solid curve shown in Fig. \ref{fig:ivsmstrength} 
corresponds to the strength for which only $2\hbar\omega$ and higher excitations are included in the configuration space.
\begin{table}[htbp]
\caption{ \label{tab:centroids} Energy centroids of the isovector spin monopole strength in the $\beta^{-}$ and the $\beta^{+}$ channel, for the nuclei $^{48}$Ca, $^{90}$Zr and $^{208}$Pb. The values calculated with the relativistic functional DD-ME2 are compared to those obtained with the Skyrme functionals SGII and SIII 
\cite{Hamamoto2000}.}
\begin{ruledtabular}
\begin{tabular}{c|cccccc}
 & \multicolumn{2}{c}{DD-ME2} & \multicolumn{2}{c}{SGII} & \multicolumn{2}{c}{SIII} \\
 & $\bar{E}_{-}$ [MeV] & $\bar{E}_{+}$ [MeV] & $\bar{E}_{-}$ [MeV] & $\bar{E}_{+}$ [MeV] & $\bar{E}_{-}$ [MeV] & $\bar{E}_{+}$ [MeV] \\
\hline
$^{48}$Ca & 34.1 & 33.9 & 35.7 & 29.6 & 35.2 & 31.5 \\
$^{90}$Zr & 40.0 & 24.2 & 40.0 & 20.8 & 39.6 & 22.1 \\
$^{208}$Pb & 37.4 & 18.3 & 39.9 & 14.3 & 38.3 & 16.5 \\
\end{tabular}
\end{ruledtabular}
\end{table}
In Table \ref{tab:centroids} we compare the energy centroids of the IVSM distributions in the  $\beta^{-}$ and the 
$\beta^{+}$ channels (excluding $0\hbar\omega$ configurations), with the corresponding values obtained 
 with two Skyrme functionals: SGII and SIII \cite{Hamamoto2000}. The agreement is very good in the 
 $\beta^{-}$ channel, whereas in the  $\beta^{+}$ channel the centroids calculated with the relativistic 
 functional DD-ME2 are found to be few MeV higher than those predicted by the two Skyrme functionals. 
 Both calculations predict a decreasing of the energy centroids with increasing mass, in agreement with 
 the results of Ref.~\cite{Auerbach1984a}.
 
\begin{figure}
\centerline{ 
  \includegraphics[scale=0.65]{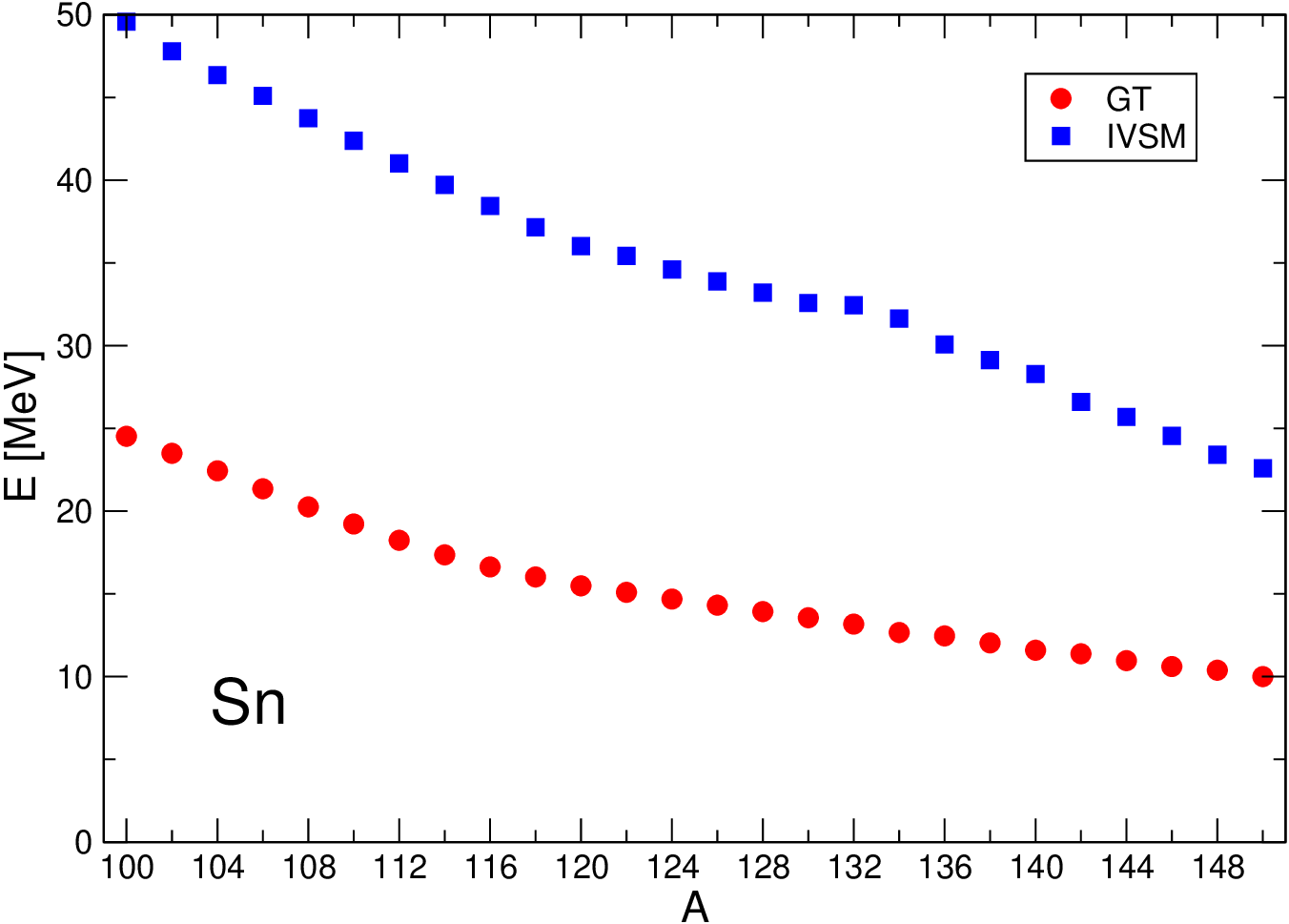}%
}
\caption{\label{fig:ivsmcentroids} (color online) Energy centroids of the isovector spin monopole strength in tin isotopes (excluding the $0\hbar\omega$ transitions), in comparison with the corresponding Gamow-Teller centroids.}
\end{figure}

 For the Sn isotopic chain with neutron-to-proton ratio ranging from $N/Z=1$ to $N/Z=2$, in Fig.
 \ref{fig:ivsmcentroids} we compare the centroids of the Gamow-Teller and the IVSM strength, considering only the $2\hbar\omega$ and higher transitions for the latter. It appears that the IVSM 
 strength function has a somewhat more pronounced mass and/or isospin dependence. In the lighter 
 Sn isotopes the IVSM centroids are found just below 50 MeV, or approximately 25 MeV above the GT 
 centroids. At mass 150 the IVSM energy centroid rapidly approaches 20 MeV, only 13 MeV 
 above the corresponding GT centroid.

\subsection{ \label{sec:gt} The L=0 strength}

The $L=0$ strength obtained in charge-exchange reactions corresponds to the squared matrix element of the $L=0$ operator
\begin{equation} \label{eq:lzero_operator}
\hat{T}_{(\pm)} = j_{L=0}(qr) \boldsymbol{\Sigma} \tau_{\pm}
\end{equation}
where $q$ is the momentum transfer. In the long wavelength limit (i.e. $q\to 0$) the spherical Bessel function can be approximated by
\begin{equation} \label{eq:sphbesselexpansion}
j_{0}(qr) \approx 1 - \frac{q^{2} r^{2}}{6} + \cdots\;,
\end{equation}
and usually only the first term is retained. However, if the momentum transfer is not negligible then,
together with the Gamow-Teller operator, the isovector spin monopole term has to be taken into account~\cite{Auerbach1984a}
\begin{equation} \label{eq:ivsmoperator}
\hat{O}_{(\pm)} = \boldsymbol{\Sigma}\tau_{\pm} - \frac{q^{2}}{6} r^{2} \boldsymbol{\Sigma}\tau_{\pm}.
\end{equation}
For a $(p,n)$ reaction the total energy and momentum of a proton with kinetic energy $T$ 
\begin{equation} \label{eq:mom_proton}
E_{p} = T + m_{p}, \qquad p_{p} = \sqrt{E_{p}^{2} - m_{p}^{2}},
\end{equation}
and for the outgoing neutron:
\begin{equation}
E_{n} = E_{p} - E_{x}, \qquad p_{n} = \sqrt{E_{n}^{2} - m_{n}^{2}},
\end{equation}
where $E_{x}$ is the excitation energy of the nucleus with respect to the ground state of the target (parent) nucleus. The momentum transfer is defined as
\begin{equation}
\left| \boldsymbol{q} \right|=  \left| \boldsymbol{p}_{p} - \boldsymbol{p}_{n} \right| = \sqrt{p_{p}^{2} + p_{n}^{2} - 2 p_{p} p_{n} \cos \vartheta},
\end{equation}
where $\vartheta$ denotes the angle between the momenta of the incoming and outgoing particles. Assuming the cross section is measured at forward angles, one can set $\vartheta \approx 0^{\circ}$, and obtain a simple expression for the momentum transfer:
\begin{equation} \label{eq:mtransferfinal}
q = \left| p_{p} - p_{n} \right|.
\end{equation}
\begin{figure}
\centerline{ 
  \includegraphics[scale=0.65]{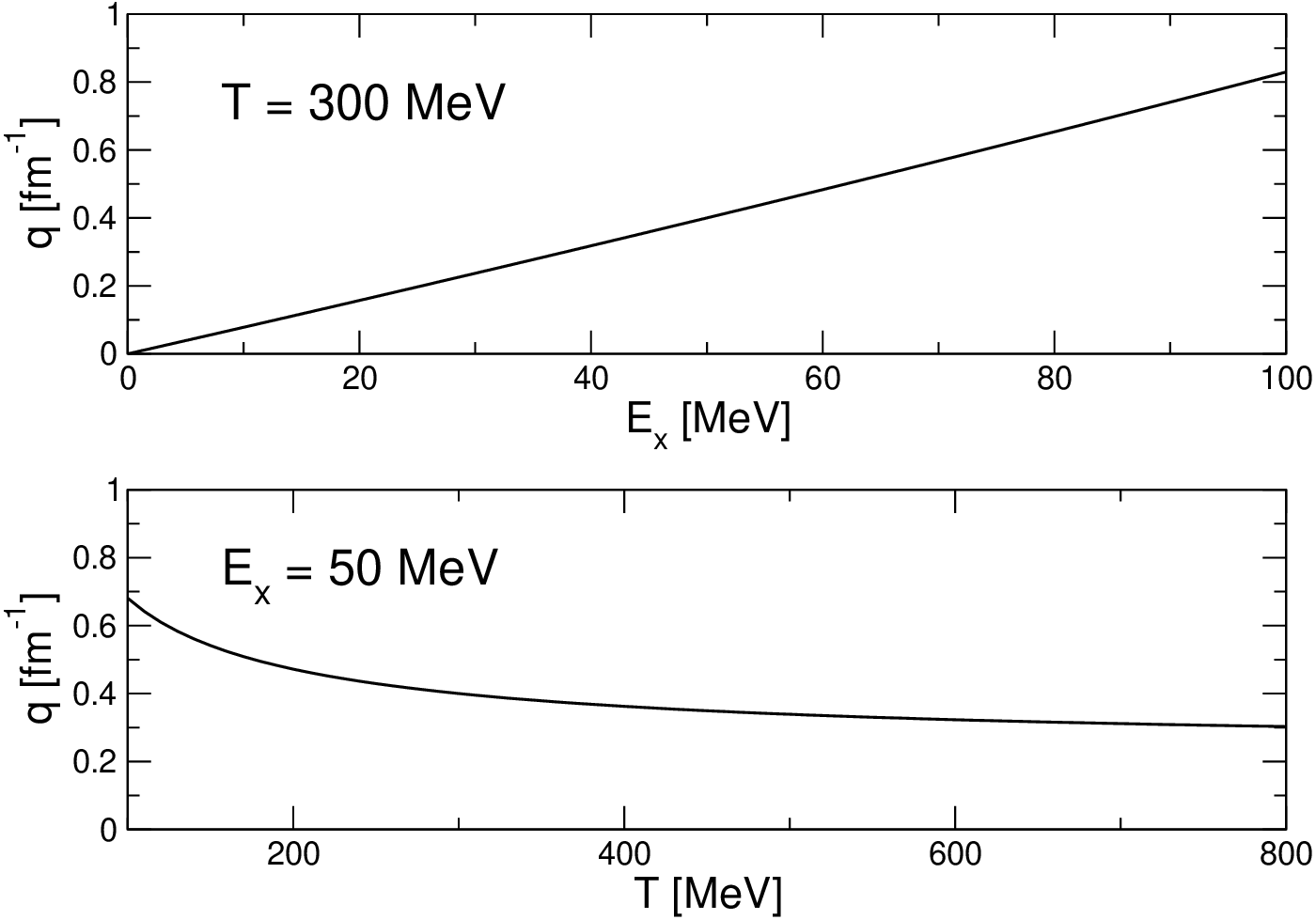}%
}
\caption{\label{fig:momentumtransfer} Momentum transfer for a $(p,n)$ reaction, calculated using Eqs. (\ref{eq:mom_proton}) - (\ref{eq:mtransferfinal}). In the upper panel $q$ is plotted for a constant kinetic energy of the incoming proton: $T = 300$ MeV, and in the lower panel for constant excitation energy of the nucleus of 
$E_{x} = 50$ MeV.}
\end{figure}
Using Eqs. (\ref{eq:mom_proton}) - (\ref{eq:mtransferfinal}), one notices that the momentum transfer depends linearly on the excitation energy of the nucleus, and has a $1/(1+T/m)^{2}$ dependence on the kinetic energy of the incoming proton, as shown in Fig.~\ref{fig:momentumtransfer}. This makes the effect of higher-order terms in the expansion Eq. (\ref{eq:sphbesselexpansion}) more pronounced for higher excitation energies and lower incoming energies. 

\begin{figure}
\centerline{ 
  \includegraphics[scale=0.65]{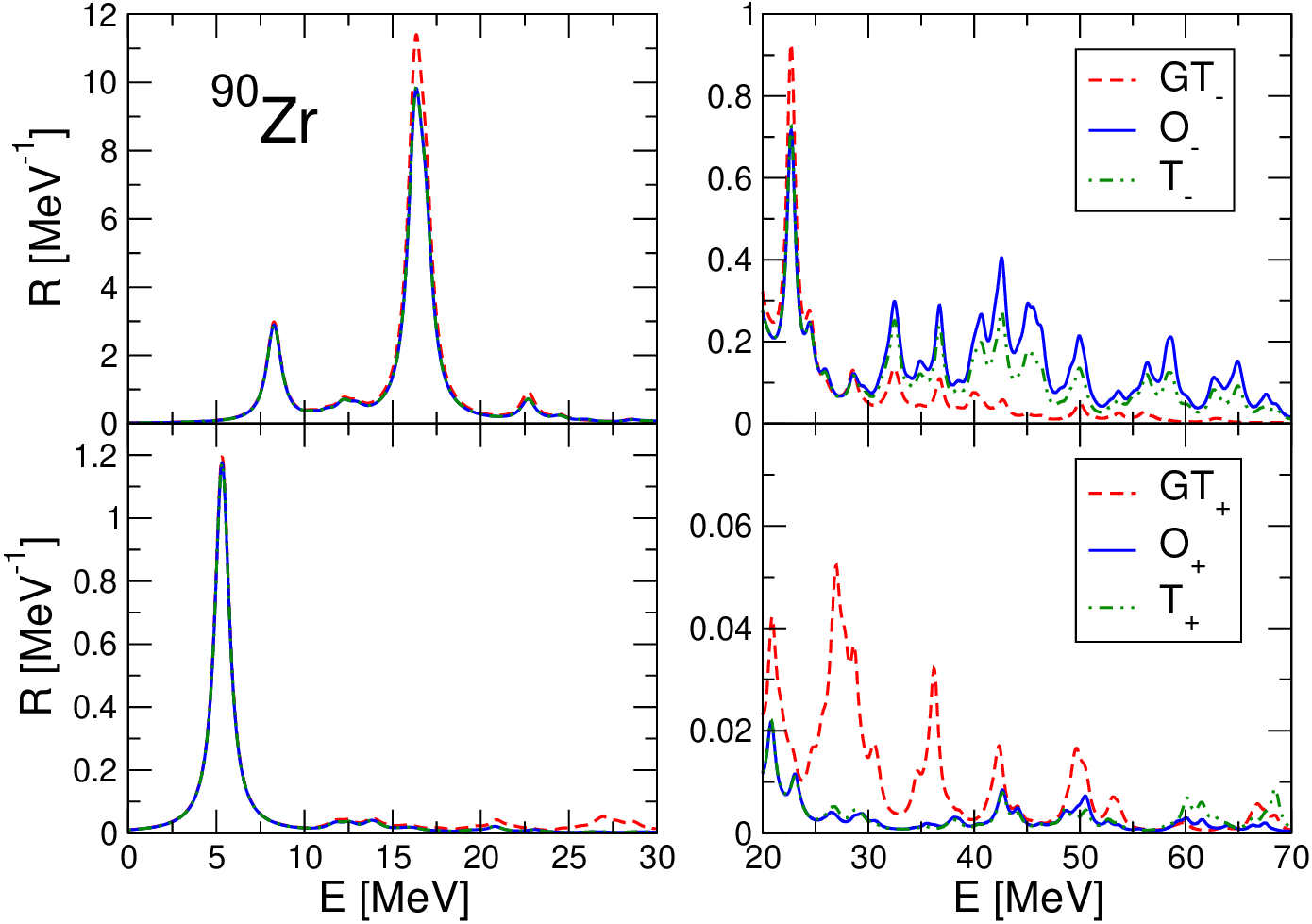}%
}
\caption{\label{fig:ivsmzr90} (color online) Comparison of the pn-RQRPA strengths obtained with the Gamow-Teller operator (dashed), the GT + IVSM operator Eq.~(\ref{eq:ivsmoperator}) (full), and the full $L=0$ operator Eq.~(\ref{eq:lzero_operator}) (dash-dotted) in $^{90}$Zr. The upper panels display the strength in the $\beta^{-}$ channel, whereas the strength in the $\beta^{+}$ channel are shown in the lower panels. Different scales are used for the region of low excitation energy below 30 MeV (left), and excitation energies in the interval 20 to 70 MeV (right).}
\end{figure}
Fig. \ref{fig:ivsmzr90} shows a comparison between the $L=0$ strengths in $^{90}$Zr,
calculated with the full operator Eq.~(\ref{eq:lzero_operator}), the 
Gamow-Teller operator, and the $q^2$-order operator Eq.~(\ref{eq:ivsmoperator}). In the upper panels we 
display the strength distributions in the $\beta^{-}$ channel. At excitation energies below 30 MeV, shown in the left panel, momentum transfer is rather small. The only significant contribution to the total strength below the GT resonance comes from the Gamow-Teller term, while in the region of the resonance the IVSM term in the operator actually reduces the strength calculated with the Gamow-Teller operator. 

The largest contribution to the strength of the $0\hbar\omega$ part of the GT + IVSM operator comes from the orbits around the Fermi surface. 
In the harmonic oscillator basis the mean value of the $r^{2}$ is equal for all orbits in a major shell, hence the following proportionality relation is obtained:
\begin{equation} \label{eq:rsquare}
\hat{O}(0\hbar\omega) = \sum_{i,j} \left\langle i \left| \left(1 - \frac{q^{2} r^{2}}{6} \right) \boldsymbol{\Sigma}\tau_{\pm} \right| j \right\rangle a^{+}_{i} a_{j} = \left(1 - \frac{q^{2}}{6} \left\langle r^{2} \right\rangle \right) \sum_{i,j} \left\langle i \left| \boldsymbol{\Sigma}\tau_{\pm} \right| j \right\rangle a^{+}_{i} a_{j},
\end{equation}
where the $\left\langle r^{2} \right\rangle$ denotes the mean value of the $r^{2}$ operator in the major shell. This relation implies that the value of the GT + IVSM matrix element will always be lower than in the case of the Gamow-Teller operator. The reduction will be greater with increasing momentum transfer, i.e. with increasing excitation energy with respect to the ground state of the parent nucleus (see Fig. \ref{fig:ivsmzr90}).

As already shown in Fig. \ref{fig:ivsmstrength}, the strength at high excitation energies originates from $2\hbar\omega$ transitions. For instance, the peak at 32.5 MeV predominantly corresponds to the $\nu 1f_{5/2} \to \pi 2f_{7/2}$ transition. The matrix elements of the GT operator are small in this case, and the IVSM term of the operator dominates. In the expansion of the spherical Bessel function $j_{0}(qr)$ successive terms have alternating signs, so that the next term reduces the strength of the isovector spin monopole mode. This is particularly visible above 40 MeV excitation energy as the next term is proportional to $q^{4}/120$. 

The corresponding strength in the $\beta^{+}$ channel is plotted in the lower panel of Fig. \ref{fig:ivsmzr90}. In the low-energy region the behavior is similar to that of the $\beta^{-}$ channel, the contribution of terms with finite momentum transfer slightly reduces the strength of the GT resonance. However, this reduction is smaller due to the lower momentum transfer involved, corresponding to the lower excitation energy. At excitation energies above 20 MeV, the strength is strongly suppressed by higher-order terms in expansion in Eq. (\ref{eq:sphbesselexpansion}), in contrast with the $\beta^{-}$ channel. Thus instead of shifting the strength to higher energies, in the $\beta^{+}$ channel finite momentum transfer simply reduces the total strength. The enhancement and reduction of the strength in the $\beta^{-}$ and $\beta^{+}$ channels, respectively, was observed in the nuclei studied in this work.

\begin{figure}
\centerline{ 
  \includegraphics[scale=0.65]{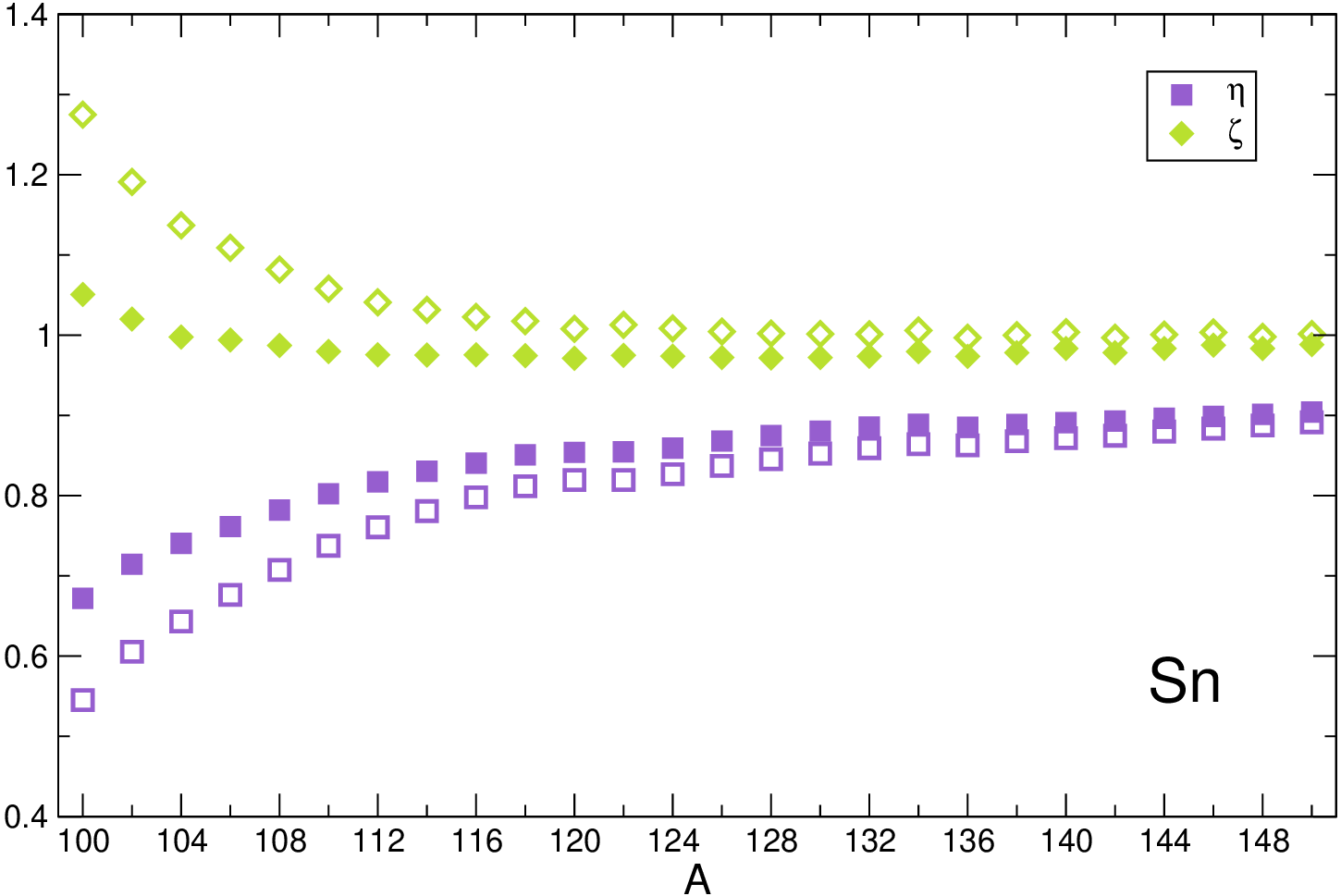}%
}
\caption{\label{fig:ratios} (color online) Ratios $\eta$ and $\zeta$ defined in Eqs. (\ref{eq:ratio_eta}) and (\ref{eq:ratio_zeta}), respectively, for the Sn isotopes. Open symbols denote ratios calculated with the GT+IVSM operator Eq.~(\ref{eq:ivsmoperator}), filled symbols are for those obtained using the full $L=0$ 
operator Eq.~(\ref{eq:lzero_operator}).}
\end{figure}

The shift of the strength to higher energies is further analyzed for the Sn isotopic chain and illustrated in Fig. \ref{fig:ratios}, where we show the effect of finite momentum transfer on the strength distribution and on the total strength in the $\beta^{-}$ channel. The energy $E_{95\%}$ below which one finds 95\% of the calculated Gamow-Teller strength, is defined by the relation
\begin{equation} \label{eq:e95}
0.95 = \frac{\sum_{i}^{E_{i} \leq E_{95\%}(N,Z)} B_{i}(GT)}{\sum_{i} B_{i}(GT)}.
\end{equation}
$E_{95\%}$ ranges from $ > 30$ MeV in the lightest tin isotopes, to $\approx 13$ MeV in $^{150}$Sn. For each isotope we calculate the ratio of the total strength (including the effect of finite momentum transfer)  below $E_{95\%}$ and the total strength,
\begin{equation} \label{eq:ratio_eta}
\eta = \frac{\sum_{i}^{E_{i} \leq E_{95\%}} B_{i} (X)}{\sum_{i} B_{i} (X)}\; ,
\end{equation} 
where $X$ denotes the operators $\hat{O}$ and $\hat{T}$ introduced in Eqs. (\ref{eq:ivsmoperator}) and (\ref{eq:lzero_operator}), respectively.
Values of $\eta<0.95$ indicate that the higher-order terms in the operator shift a portion of the strength from the resonance to higher excitation energies. To see the effect of momentum transfer on the total strength, we also plot  the ratio of the total strength calculated with the operators $O$ Eq.~(\ref{eq:ivsmoperator}) and $T$ Eq.~(\ref{eq:lzero_operator}), and the total Gamow-Teller strength
\begin{equation} \label{eq:ratio_zeta}
\zeta = \frac{\sum_{i} B_{i} (X)}{\sum_{i} B_{i} (GT)}.
\end{equation}

\begin{figure}
\centerline{ 
  \includegraphics[scale=0.65]{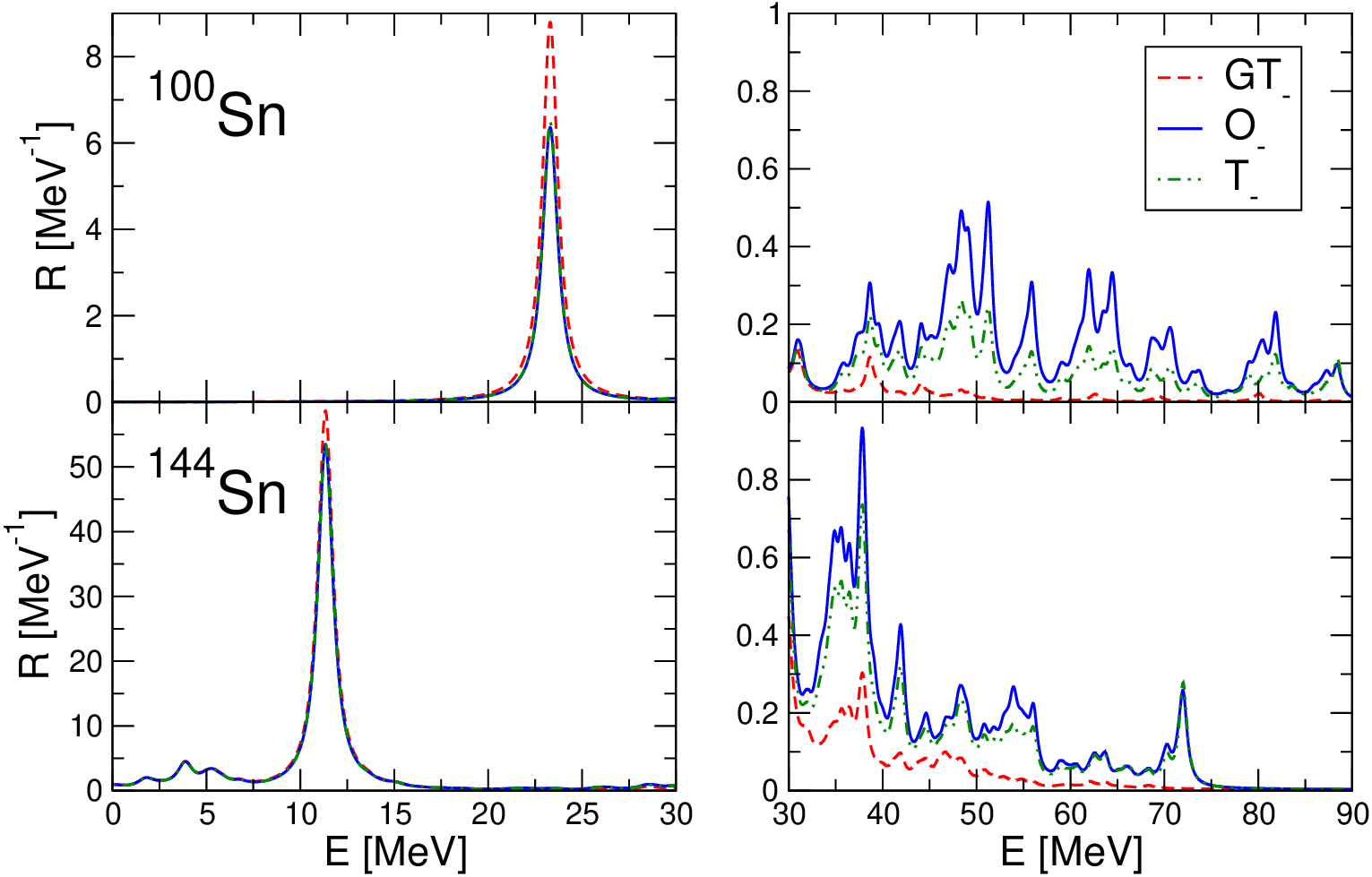}%
}
\caption{\label{fig:tincomparison} (color online) Comparison of the $\beta^{-}$ strength distributions in $^{100}$Sn (upper panels) and $^{144}$Sn (lower panels), calculated with the Gamow-Teller operator, the GT+IVSM operator Eq.~(\ref{eq:ivsmoperator}) and the full $L=0$ operator Eq.~(\ref{eq:lzero_operator}). Note the different scales 
used for the resonance region and the region of high-excitation energies.}
\end{figure}

For Sn isotopes with masses in the interval $100 \le A \le 150$ the ratios defined in Eqs. (\ref{eq:ratio_eta}) and (\ref{eq:ratio_zeta}), are plotted in Fig. \ref{fig:ratios}. Open symbols denote results obtained with the GT+IVSM operator Eq.~(\ref{eq:ivsmoperator}). One notices that for light isotopes a rather large amount of strength is found above the Gamow-Teller resonance. The fraction of strength found at lower energies increases with the addition of neutrons to $\eta = 0.8$ for $^{116}$Sn. From $A=116$ the ratio $\eta$ is a linear function of the mass, and reaches the value of $\eta = 0.9$ for $^{150}$Sn. The ratio of the total strengths $\zeta$ is considerably larger than 1 for lighter isotopes, but rapidly converges to $\zeta = 1$ with the addition of neutrons.
With filled symbols we denote results obtained using the full $L=0$ operator Eq. (\ref{eq:lzero_operator}). The prominent feature is that, with respect to the GT+IVSM operator, the ratio $\eta$ increases and $\zeta$ decreases for all isotopes. Since the next term in the expansion Eq. (\ref{eq:sphbesselexpansion}) is proportional to $q^{4}$, a relatively large momentum transfer is necessary for an effect to be noticeable (see also the right panels in Figs. \ref{fig:ivsmzr90} and \ref{fig:tincomparison}). Therefore, this term in the expansion does not affect the resonance but reduces the strength at high energies, and in this way reduces the total strength and increases the fraction of the strength below the GT resonance. Because the energy centroid of the IVSM is higher in isotopes with a low number of excess neutrons, with correspondingly large momentum transfer, higher-order terms have a more pronounced effect in Sn isotopes with $A \approx 100$. The largest differences of the ratios $\eta$ and $\zeta$ with respect to those calculated with the GT+IVSM operator Eq.~(\ref{eq:ivsmoperator}), are found in the lightest Sn nuclei. 

\begin{figure}
\centerline{ 
  \includegraphics[scale=0.65]{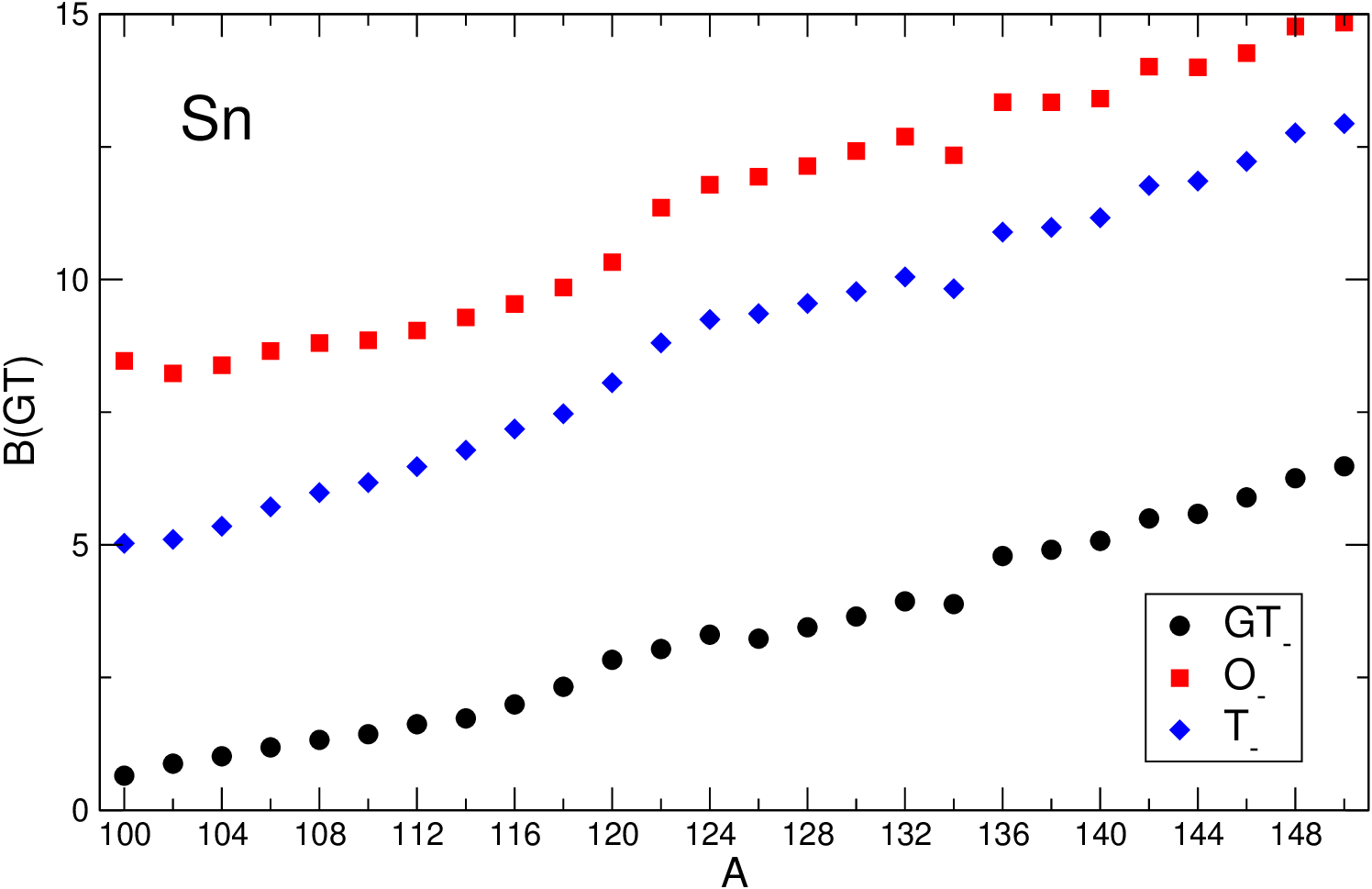}%
}
\caption{\label{fig:hestrength} (color online) $\beta^{-}$ pn-RQRPA strength located at energies above $E_{95\%}$ in Sn isotopes with $100 \le A \le 150$. The GT operator (circles), the GT + IVSM operator defined in Eq. (\ref{eq:ivsmoperator}) (squares), and the full $L=0$ operator Eq. (\ref{eq:lzero_operator}) (diamonds), have been used in the calculation of the strength distributions.}
\end{figure}

This effect is further illustrated in Fig.~\ref{fig:tincomparison} where we compare the strength calculated using the Gamow-Teller operator, with those obtained using the GT+IVSM operator Eq.~(\ref{eq:ivsmoperator}) and the 
full $L=0$ operator Eq.~(\ref{eq:lzero_operator}), in two Sn isotopes: $^{100}$Sn and $^{144}$Sn. In the former the resonance is at 23 MeV excitation energy with respect to the ground state of the parent nucleus, and the corresponding momentum transfer is $q=0.181$ fm$^{-1}$ for the incoming proton kinetic energy $T=300$ MeV. The inclusion of the IVSM term reduces the strength of the resonance by approximately 30\%. In $^{144}$Sn the resonance is at 11 MeV, with the corresponding momentum transfer $q=0.086$ fm$^{-1}$. Since the square of momentum transfer appears in the IVSM operator, in this case the effect on the resonance is significantly smaller. It is important to note that, even though the relative reduction of the resonance is more pronounced in the lighter isotope, more strength is actually subtracted from the resonance in $^{144}$Sn: $B(GT_{-}) - B(T_{-}) = 8.0$, than in $^{100}$Sn: $B(GT_{-}) - B(T_{-}) = 3.7$. In Fig. \ref{fig:hestrength} we plot the strength that is calculated at energies above $E_{95\%}$ for the Sn isotopic chain. One notices that the difference between the strength 
obtained using the operators defined in Eqs. (\ref{eq:lzero_operator}) or (\ref{eq:ivsmoperator}), and the GT strength is practically constant. Because of this in light isotopes there is enough additional strength to overcome the reduction of the resonance and even increase the total strength. In heavy isotopes the strength at high energies mostly compensates for the strength lost in the resonance but does not increase the total strength. 

In Table \ref{tab:fullratios} we display the values of the ratio $\zeta$ defined in Eq.~(\ref{eq:ratio_zeta}), in both the $\beta^{-}$ and the $\beta^{+}$ channel for three representative nuclei: $^{48}$Ca, $^{90}$Zr and $^{208}$Pb. The momentum transfer corresponds to a kinetic energy of 300 MeV for the incoming proton (neutron). The results for $^{90}$Zr are particularly important because of the recent analysis of both $(p,n)$ and $(n,p)$ data \cite{Wakasa1997,Yako2005}, that determined the GT quenching factor 
\begin{equation}
Q \equiv {{S_{\beta^{-}}^{\text{GT}} - S_{\beta^{+}}^{\text{GT}}}\over{3(N-Z)}} = 0.88 \pm 0.06 \;.
\end{equation}
In the $\beta^{-}$ channel the $L=0$ strength was measured up to 50 MeV excitation energy: 
$S_{\beta^{-}}^{L=0} = 33.5 \pm 0.6 (\text{stat.}) \pm 0.4 (\text{MD}) \pm 4.7 (\hat{\sigma}_{GT})$. 
Employing a distorted wave impulse approximation (DWIA) model to estimate the contribution of the isovector spin monopole strength, the assumption was made that the complete IVSM strength is concentrated in a single state at 35 MeV. The estimated contribution was then subtracted from the measured strength, and the value of the total Gamow-Teller strength was determined: $S_{\beta^{-}}^{\text{GT}} = 29.3 \pm 0.5 (\text{stat.}) \pm 0.4 (\text{MD}) \pm 0.9 (\text{IVSM}) \pm 4.7 (\hat{\sigma}_{GT})$. This means that the IVSM contribution enhances the total strength by approximately 15\%. Our results using the full $L=0$ operator show (cf. Table \ref{tab:fullratios}), that the total strength in the $\beta^{-}$ channel is not modified by the inclusion of higher order terms, and their only effect 
is to shift part of the strength to energies above the resonance. Therefore, the result of the present calculation 
implies $S_{\beta^{-}}^{\text{GT}} = S_{\beta^{-}}^{L=0}$.

The experimental value of the $L=0$ strength in the $\beta^{+}$ channel was determined in Ref.~\cite{Yako2005}: $S_{\beta^{+}}^{L=0} = 5.4 \pm 0.4 (\text{stat.}) \pm 0.3 (\text{MD}) \pm 0.9 (\hat{\sigma}_{GT})$. After subtracting the IVSM strength, the value of the GT$^{+}$ strength was deduced: $S_{\beta^{+}}^{\text{GT}} = 2.9 \pm 0.4 (\text{stat.}) \pm 0.3 (\text{MD}) \pm 0.3 (\text{IVSM}) \pm 0.5 (\hat{\sigma}_{GT})$. The present calculation indicates, however, that the total GT strength in the $\beta^{+}$ channel is actually reduced by $\approx 15\%$ by the inclusion of finite momentum-transfer terms. Therefore, using the total measured $L=0$ strength, we deduce the GT strength in the $\beta^{+}$ channel: $S_{\beta^{+}}^{\text{GT}}=6.3$. The deduced value for the Ikeda sum rule: $S_{\beta^{-}}^{\text{GT}} - S_{\beta^{+}}^{\text{GT}} = 33.5 - 6.3 = 27.2$, is consistent with the quenching factor extracted from  data in Ref.~\cite{Yako2005}. However, in the $\beta^{-}$ channel data were only obtained below 50 MeV excitation energy, whereas our calculation predicts that approximately 6\% of the total strength is located above this energy. Assuming that the measured strength actually corresponds to only 94\% of the total strength, we obtain $ S_{\beta^{-}}^{\text{GT}}= 35.6$ and, therefore, the value of the sum rule: $ S_{\beta^{-}}^{\text{GT}} - S_{\beta^{+}}^{\text{GT}} = 29.3$. Considering the experimental uncertainty, in particular the one originating from the Gamow-Teller unit cross section, this result may indicate that no quenching of the experimental strength with respect to the Ikeda sum rule occurs. We note that arguments for this conclusion were already put forward from the point of view of the shell model~\cite{Caurier1995}.
\begin{table}[hbtp]
\caption{ \label{tab:fullratios} The ratio $\zeta$ defined in Eq. (\ref{eq:ratio_zeta}), for the $\beta^{-}$ and 
$\beta^{+}$ channels in $^{48}$Ca, $^{90}$Zr and $^{208}$Pb. The second and third columns display results 
calculated with the GT+IVSM operator Eq.~(\ref{eq:ivsmoperator}). In the last two columns we show the results obtained with the full $L=0$ operator Eq.~(\ref{eq:lzero_operator}). The momentum transfer corresponds to a kinetic energy of 300 MeV for the incoming nucleon.}
\begin{ruledtabular}
\begin{tabular}{c|cccc}
 & \multicolumn{2}{c}{$\left(1 - \frac{q^{2}r^{2}}{6}\right)\boldsymbol{\Sigma}\tau_{\pm}$} & \multicolumn{2}{c}{$j_{0}(qr) \boldsymbol{\Sigma}\tau_{\pm}$} \\
 & $\beta^{-}$ & $\beta^{+}$ & $\beta^{-}$ & $\beta^{+}$ \\
\hline
$^{48}$Ca & 1.043 & 0.821 & 1.030 & 0.661 \\
$^{90}$Zr & 1.043 & 0.871 & 0.999 & 0.851 \\
$^{208}$Pb & 0.952 & 0.810 & 0.877 & 0.342 \\
\end{tabular}
\end{ruledtabular}
\end{table}

\section{ \label{sec:conclusion} Conclusion and outlook}

An accurate determination of Gamow-Teller strength remains a challenge for  charge-exchange reaction experiments. With the progress of experiments that can provide data on the nuclear response at high excitation energies, the effect of finite momentum transfer must be taken into account. In this work the $L=0$ strength has been analyzed in the Sn isotopic chain, $^{48}$Ca, $^{90}$Zr and $^{208}$Pb. Employing the RHB + pn-RQRPA framework, we have compared strength functions calculated using the GT operator, the GT plus isovector spin monopole mode term, and the operator that contains the full momentum transfer dependence.

The transition strength for the pure isovector spin monopole operator has been calculated for $^{48}$Ca, $^{90}$Zr and $^{208}$Pb. We have decomposed the contributions to the strength into $0\hbar\omega$, and $2\hbar\omega$ and higher components and found that the $0\hbar\omega$ contributes mainly to the resonance. The large number of $2\hbar \omega$ and higher transitions form a very broad structure at excitation energies between 30 and 60 MeV. The calculated energy centroids are in very good agreement with values previously obtained using two different Skyrme interactions. The dependence of the centroids on neutron number has been shown for the Sn isotopic chain with the neutron-to-proton ratio in the interval from $N/Z=1$ to $N/Z=2$. The IVSM centroids are located at high excitation energies, ranging from 25 MeV above the GT centroids for the lightest isotopes, to 12 MeV for the heaviest. 

Evaluations of GT strength from experimental cross sections of charge-exchange reactions take into account the isovector spin monopole mode but, because of the unknown distribution of the IVSM strength, its contribution is subtracted incoherently from the total measured strength. To analyze the validity of this procedure, we have calculated the $L=0$ strength using the GT+IVSM operator Eq.~(\ref{eq:ivsmoperator}) and the full $L=0$ operator Eq.~(\ref{eq:lzero_operator}). It has been found that the inclusion of the isovector spin monopole term
contributes to the strength at high excitation energies, and also reduces the strength of the resonance. The shift of the strength to higher excitation energies has been analyzed for the Sn isotopic chain with masses in the range $100<A<150$. The total $L=0$ strength for isotopes with low number of excess neutrons is enhanced, whereas it is not modified for isotopes with $A \ge 120$. The full $L=0$ operator only changes the strength at high excitation energies, i.e. for large momentum transfer. A similar analysis of the effect of finite momentum transfer has been  performed for $^{48}$Ca, $^{90}$Zr and $^{208}$Pb. For $^{90}$Zr, in particular, the results have been compared with a recent analysis of the GT quenching factor based on $(p,n)$ and $(n,p)$ data. We have found that the total strength in the $\beta^{-}$ channel is not modified by the inclusion of higher order terms, i.e. $S_{\beta^{-}}^{\text{GT}} = S_{\beta^{-}}^{L=0}$. The strength in the $\beta_{+}$ channel is reduced by approximately 15\% by the inclusion of finite momentum-transfer terms, contrary to the assumption made in the analysis of experimental cross sections. Combining these results with the model prediction that 6\% of the strength in the $\beta^{-}$ is located above 50 MeV excitation energy, we find that the Ikeda sum rule is satisfied within experimental uncertainty. 

The determination of Gamow-Teller strength, complicated however by the excitation of the IVSM mode, was also performed using the ($^{3}$He,$t$) reaction on $^{208}$Pb~\cite{Zegers2003} and $^{150}$Nd~\cite{Guess2011}. Because the IVSM transition density has a node close to the surface, probes that penetrate deep into the nucleus display smaller cross sections due to the cancellation of contributions from the surface and the bulk. In contrast, probes absorbed at the surface have larger cross sections because there is no contribution from the volume region. An analogous effect can be obtained using probes with different energies (see Sec. IV.A in Ref.~\cite{Prout2000}). The present calculation does not differentiate between various probes, and the only effect of the energy of the incoming probe is on momentum transfer. It would be interesting to perform a study of the interference of Gamow-Teller and isovector spin-monopole modes, taking into account the characteristics of the experimental  probe. One could, in particular, combine the RQRPA transition densities with a DWIA calculation, and compare the resulting cross sections with the experiment.

\begin{acknowledgments}
This work was supported in part by the Helmholtz International Center for FAIR within the framework of the LOEWE program launched by the State of Hesse, by the Deutsche Forschungsgemeinschaft through contract SFB~634 and by the MZOS - project 1191005-1010. We would like to thank R. G. T. Zegers for helpful comments and suggestions. 
\end{acknowledgments}

\end{document}